\documentclass[aps,twocolumn,amssymb,showpacs]{revtex4}
\usepackage{epsfig}
\usepackage{bm}
\begin{document}
\title{Tailoring Phase Space : A Way to Control Hamiltonian Transport}
\author{Guido Ciraolo}
\affiliation{Facolt\`a di Ingegneria, Universit\`a di Firenze,
via S. Marta, I-50129 Firenze, Italy, and I.N.F.M. UdR Firenze}
\author{Cristel Chandre}
\affiliation{CPT-CNRS, Luminy Case 907, F-13288 Marseille Cedex 9,
France}
\author{Ricardo Lima}
\affiliation{CPT-CNRS, Luminy Case 907, F-13288 Marseille Cedex 9,
France}
\author{Michel Vittot}
\affiliation{CPT-CNRS, Luminy Case 907, F-13288 Marseille Cedex 9,
France}
\author{Marco Pettini}
\affiliation{Istituto Nazionale di Astrofisica, Osservatorio
Astrofisico di Arcetri, Largo Enrico Fermi 5, I-50125 Firenze, Italy,
and I.N.F.M. UdR Firenze and I.N.F.N. Sezione di Firenze}
\author{Philippe Ghendrih}
\affiliation{Association Euratom-CEA, DRFC/DSM/CEA, CEA Cadarache,
F-13108 St. Paul-lez-Durance Cedex, France}
\date{\today}

\begin{abstract}
We present a method to control transport in Hamiltonian systems. We provide an algorithm 
- based on a perturbation of the original Hamiltonian {\em localized in phase space} - to design {\it small} control terms that are able to create isolated barriers of transport without modifying other parts of phase space.  
We apply this method of localized control to a forced pendulum model and to a system describing the motion of charged particles in a model of turbulent electric field.

\end{abstract}

\maketitle
Transport governed by chaotic motion is a main issue in nonlinear dynamics. In several contexts,
transport in Hamiltonian systems leads to undesirable effects.
For example, chaos in beams of particle accelerators leads to a weakening of the beam luminosity.
Similar problems are encountered in free electron lasers. In magnetically confined fusion plasmas,
the so called anomalous transport, which has its microscopic origin in a chaotic transport of charged
particles, represents a challenge to the attainment of high performance in fusion devices.
One way to control transport would be that of reducing or
suppressing chaos. There exist numerous attempts to cope with this problem
of controlling chaos~\cite{review}.
However, in many situations, it would be desirable to control the transport properties without significantly
altering the original setup of the system under investigation nor its overall chaotic structure.\\
\indent In this article, we address the problem of control of transport
in Hamiltonian systems. We consider the class of Hamiltonian systems that can be written
in the form $H=H_0+\epsilon V$ that is an integrable 
Hamiltonian $H_0$ (with action-angle variables)
plus a perturbation $\epsilon V$. 
For these Hamiltonians we provide a method to construct a control term $f$ of order $\epsilon^2$ with a {\it finite support} in phase space,
 such that the controlled Hamiltonian $H_c=H_0+\epsilon V+ f$ has isolated invariant tori. For Hamiltonian systems with two degrees of freedom, these invariant tori act as barriers in phase space.
 For higher dimensional systems, the barriers of transport can be formed 
as a localized collection
 of invariant tori. The idea is to slightly and locally modify the perturbation and create isolated barriers of transport without modifying the dynamics inside and outside the neighborhood of the barrier. Furthermore, we require that, in order to compute the control term, only the knowledge of the Hamiltonian inside the designed support region of control is needed.\\
The main motivations for a localized control are the following~: Very often the control is only conceivable in some specific regions of phase space (where the Hamiltonian can be known and/or an implementation is possible). 
Or, there are cases for which it is desirable to stabilize only a given region of phase space without modifying the chaotic regime outside the controlled region. This can be used to bound the motion of particles without changing the phase space on both sides of the barrier.\\
 \indent Our algorithm for a localized control contains three steps~: a global control in the framework of Refs.~\cite{michel,guido1}, a selection of the desired invariant tori to be preserved, and a localization of the control term.\\
The global control term 
$f$ of order $\epsilon^2$ we construct is
such that the controlled Hamiltonian given by 
$H_c=H_0+\epsilon V+f$ 
is integrable or close to integrable, i.e. such that $H_c$ is canonically 
conjugate to $H_0$ up to some correction terms.\\ 
\indent Let us fix a Hamiltonian $H_0$.
We define the linear operator $\{H_0\}$ by $\{H_0\}H=\{H_0,H\},$
where $\{\cdot~,\cdot\}$ is the Poisson bracket.
 The operator
\(\{{H_0}\} \) is not invertible. We consider a pseudo-inverse of \( \{{H_0}\} \), denoted by $\Gamma$, satisfying
\begin{equation}
\{{H_0}\}^{2}\ \Gamma = \{{H_0}\}.
\label{gamma}
\end{equation}
If the operator $\Gamma$ exists, it is not unique in general. For a given $\Gamma$, we define the {\em resonant} operator $\mathcal R$ as
\begin{equation}
{\mathcal R} = 1-\{H_0\}\Gamma,
\end{equation} 
We notice that Eq.~(\ref{gamma}) becomes $\{{H_0}\} \mathcal R = 0$.
A consequence
is that any element ${\mathcal R} V$ is constant under the flow of $H_0$. 
\\ \indent Let us now assume that $H_0$ is integrable with action-angle variables 
$({\bf A},\bm{\varphi})\in {\mathbb R}^l\times \mathbb{T}^l $ where ${\mathbb T}^l$ is the $l$-dimensional torus. Moreover, we assume that $H_0$ is linear in the action variables, so that $H_0={\bm \omega}\cdot {\bf A}$, where the frequency vector ${\bm \omega}$ is any vector of ${\mathbb R}^l$.
The operator $\{H_0\}$ acts on 
$
V=\sum_{{\bf k}\in {\mathbb Z}^l}V_{\bf k}({\bf A})e^{i{\bf k}\cdot{\bm\varphi}},
$
as
$
\{H_0\}V({\bf A},\bm{\varphi})=\sum_{\bf k}i{\bm \omega}\cdot{\bf k}~V_{\bf k}({\bf A})e^{i{\bf k}\cdot{\bm\varphi}}.
$
A possible choice of $\Gamma$ is then
$$
\Gamma V({\bf A},\bm{\varphi})=
\sum_{{\bf k}\in{\mathbb Z^l}\atop{{\bm\omega}\cdot{\bf k}\neq0}}
\frac{V_{\bf k}({\bf A})}
{i{\bm \omega} \cdot{\bf k}}~~e^{i{\bf k}\cdot{\bm\varphi}}.
$$
We notice that this choice of $\Gamma$ commutes with $\{H_0\}$.
\\ \indent The operator ${\mathcal R}$ is the projector on the resonant 
part of the perturbation:
\begin{equation}
{\mathcal R}V=\sum_{{\bm\omega}\cdot{\bf k}=0}
V_{\bf k}({\bf A})e^{i{\bf k}\cdot{\bm\varphi}}\label{eqn:RV}.
\end{equation}
>From the operators $\Gamma$ and ${\mathcal R}$, we construct a control term for the perturbed Hamiltonian $H_0+V$, i.e.\ we construct $f$ such that the controlled Hamiltonian $H_c=H_0+V+f$
is canonically conjugate to $H_0+\mathcal R V$. This conjugation is given by the following equation
\begin{equation}
e^{\{\Gamma V\}}(H_0+V+f)=H_0+{\mathcal R} V,
\label{prop1}
\end{equation}
where
\begin{equation}
f=\sum_{n=1}^{\infty}\frac{(-1)^n}{(n+1)!}\{\Gamma V\}^n
(n{\mathcal R}+1)V.
\label{eqn:ctf}
\end{equation}
As a consequence, if $V$ is of order $\epsilon$, the largest term in the expansion of $f$ is of order $\epsilon^2$.
We notice that $H_0$ and ${\mathcal R}V$ are two conserved quantities for the 
flow defined by $H_0+{\mathcal R}V$ (even if $H_0+{\mathcal R}V$ 
does depend on the 
angles in general). Therefore, for Hamiltonian systems with two degrees of 
freedom, $H_0+{\mathcal R}V$ as well as the controlled Hamiltonian $H_c$ are 
integrable. For higher dimensional systems, if ${\bm\omega}$ is non-resonant, 
i.e. $\forall~{\bf k}\in{\mathbb Z}^l\setminus\{{\bf 0}\}$, ${\bm\omega}\cdot{\bf k}\neq 0$,
$H_c$ is also integrable since it is canonically conjugate 
to $H_0+{\mathcal R}V$ which only depends on the actions.

Therefore the control term given by Eq.~(\ref{eqn:ctf}) is able to recreate invariant 
tori, the ones of $H_0+{\mathcal R}V$. 
Truncations of this control term are also able to recreate invariant tori~\cite{guido1}. 
Of course, the closer to integrability 
the more invariant tori are created. After the computation of the control term, the second step is to select a given region of phase space where the localized control acts. This region has to contain an invariant torus created by the previous control and also a small neighborhood of it. The invariant torus to be created can be selected by its frequency using Frequency Map Analysis~\cite{laskar}. The third step is to multiply the 
control term by a smooth window around the selected region where the control 
has to be applied~: The locally controlled Hamiltonian is 
$
H_c=H_0+V+\chi f,
$
where $\chi$ is a characteristic function of a neighborhood $P$ of the selected invariant torus i.e.\ $\chi(x)=1$ for $x\in P$, $\chi(x)\not=0$ in a small neighborhood of $P$ in order to have a smooth function, and $\chi(x)=0$ otherwise. The phase space of the controlled Hamiltonian $H_c$ is very similar to the phase space of the uncontrolled Hamiltonian (since the control term acts only locally) with the addition of the selected invariant torus. \\
The justification for the localized control follows from the KAM theorem~: The controlled Hamiltonian can be written as
$
H_c=H_0+V+f+(\chi-1)f
$.
Since $H_0+V+f$ has the selected invariant torus (by construction), also the controlled Hamiltonian $H_c$ has this invariant torus provided that the hypothesis of the KAM theorem are satisfied, namely that the perturbation $(\chi-1)f$
is sufficiently small and smooth in the neighborhood of the invariant torus.\\ 

The first application of the localized control is on the following forced 
pendulum model~\cite{esca85}
\begin{equation}
\label{eqn:fp}
H(p,x,t)=\frac{1}{2}p^2+\varepsilon \left[ \cos x+\cos(x-t)\right].
\label{Hpend}
\end{equation}
for $\varepsilon=0.034$. We notice that for $\varepsilon\geq 0.02759$ there 
are no longer any invariant rotational (KAM) tori~\cite{chanPR}.

\begin{figure}
\unitlength 1cm
\begin{picture}(7.5,6)(0,0)
\put(0,0){\epsfig{file=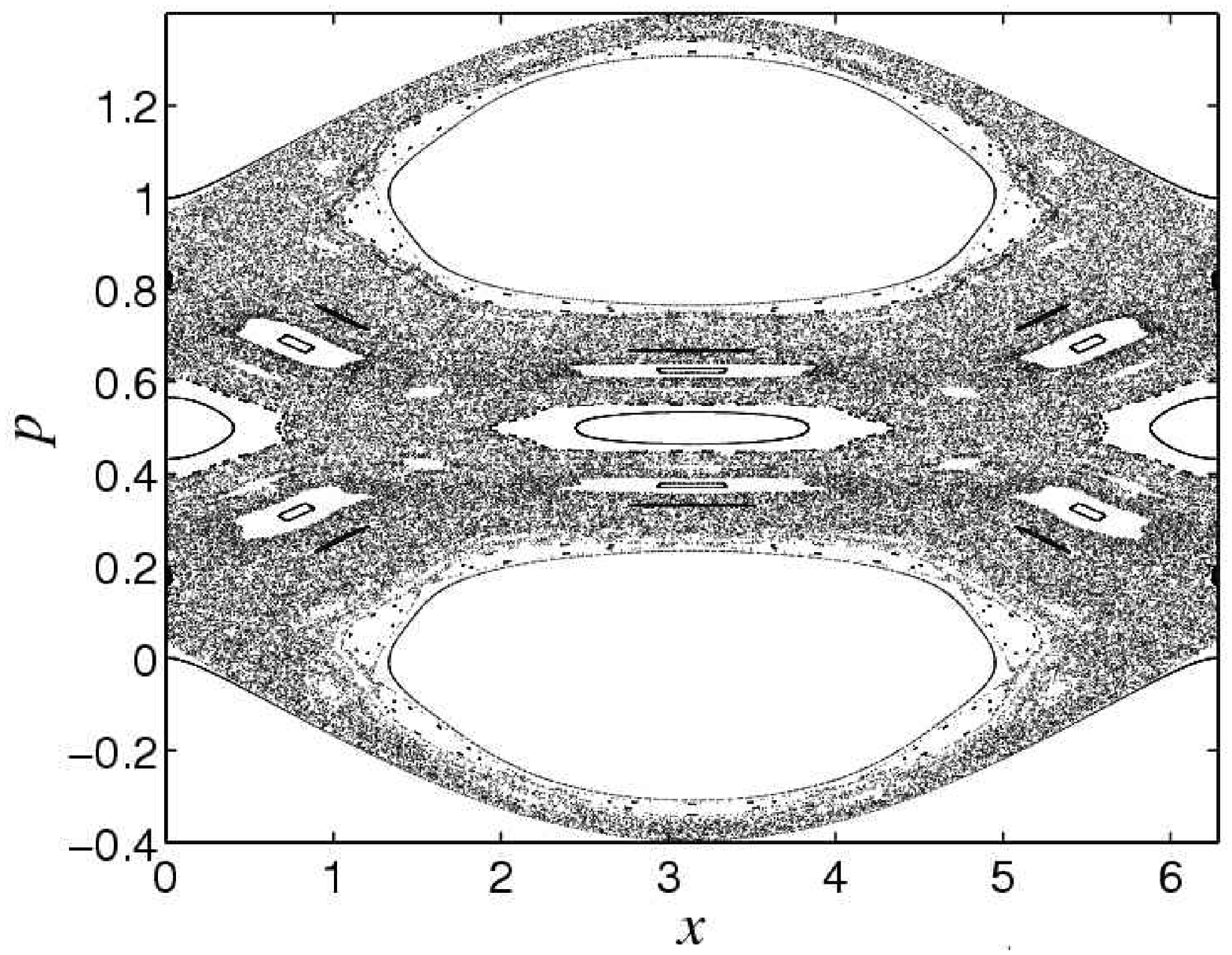,width=7.5cm,height=6.3cm}}
\put(3,3.9){\epsfig{file=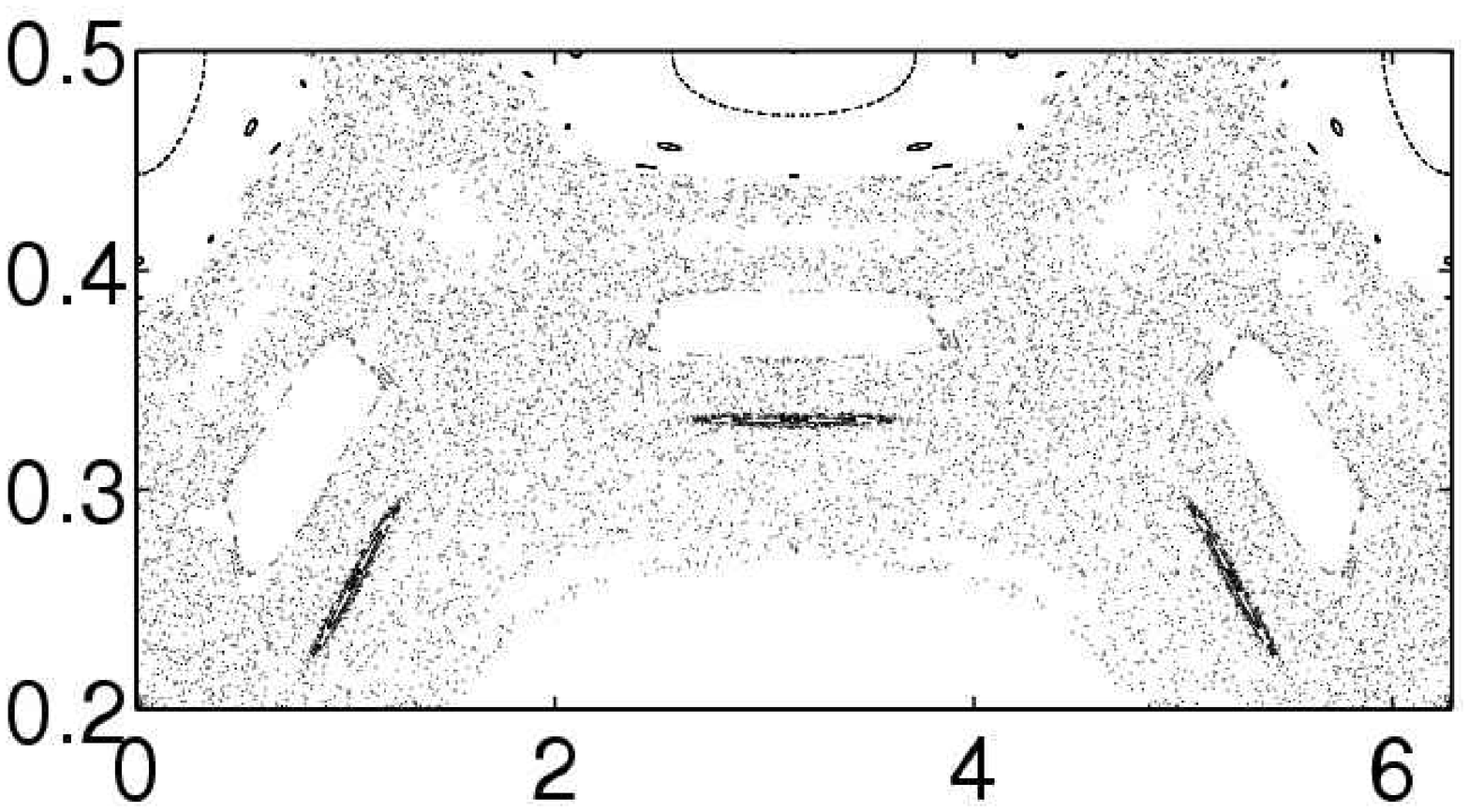,width=4.3cm,height=2.3cm}}
\end{picture}
\caption{Poincar\'e surface of section of Hamiltonian (\ref{eqn:fp}) with $\varepsilon=0.034$ (enlargement in the inset).}
\label{fig1}
\end{figure}

For the construction of the control term, we choose $H_0=E+\omega p$ and $V=\sqrt{\varepsilon} \left[ p^2/2+\cos x+\cos(x-t)\right]$ after a translation of the momentum $p$ by $\omega$, and a rescaling (zoom in phase space) by a quantity $\sqrt{\varepsilon}$.
The control term of Hamiltonian~(\ref{Hpend}) becomes
\begin{equation}
f_2(x,t)=-\frac{\varepsilon^2}{2}\left(\frac{\cos x}{\omega}+\frac{\cos(x-t)}{\omega-1}\right)^2,
\label{eqn:f2fpa}
\end{equation}
around a region near $p\approx \omega$. In the numerical implementation, we use
$\omega=(3-\sqrt 5)/2$.
We notice that the perturbation has a norm (defined as the maximum of its amplitude) of $6.8\times 10^{-2}$ whereas the control term has a norm of $2.7\times 10^{-3}$ for $\varepsilon=0.034$. The control term is small (about 4\%) compared to the perturbation. We notice that it is still possible to reduce
the amplitude of the control (by a factor larger than 2) and still get an invariant torus of the desired frequency.\\
\indent The phase space of Hamiltonian~(\ref{eqn:fp}) with the approximate control term~(\ref{eqn:f2fpa}) for $\varepsilon=0.034$ shows that a lot of invariant tori are created with the addition of the control term~\cite{guido3}. Using the renormalization-group transformation~\cite{chanPR}, we have checked the existence of the invariant torus with frequency $\omega$ for the Hamiltonian $H+f_2$ with $\varepsilon\leq 0.06965$. \\
The next step is to localize $f_2$ given by Eq.~(\ref{eqn:f2fpa}) around a chosen invariant torus created by $f_2$~: We assume that the controlled Hamiltonian~$H+f_2$ has an invariant torus with the frequency $\omega$. 
We locate this invariant torus using Frequency Map Analysis~\cite{laskar}. 
Once the invariant torus has been located, we construct an approximation of the invariant torus for the Hamiltonian $H+f_2$ of the form $p=p_0(x,t)$. Then we consider the following localized control term~:
\begin{equation}
\label{eqn:f2fpaL}
f_2^{(L)}=f_2(p,x,t)\chi(p-p_0(x,t)),
\end{equation}
where $\chi$ is a smooth function with finite support around zero. More precisely, we have chosen $\chi(p)=1$ for $\vert p\vert \leq \alpha$, $\chi(p)=0$ for $\vert p\vert \geq \beta$ and a polynomial for $\vert p\vert \in [\alpha,\beta]$ for which $\chi$ is a $C^3$ even function. The function $p_0$ and the parameters $\alpha$, $\beta$ are determined numerically ($\alpha=5\times 10^{-3}$ and $\beta=1.5\alpha$). The measure of the support 
of $\chi$ is about $1\%$ of the measure of the support 
of the global control.
\begin{figure}
\unitlength 1cm
\begin{picture}(7.5,6.3)(0,0)
\put(0,0){\epsfig{file=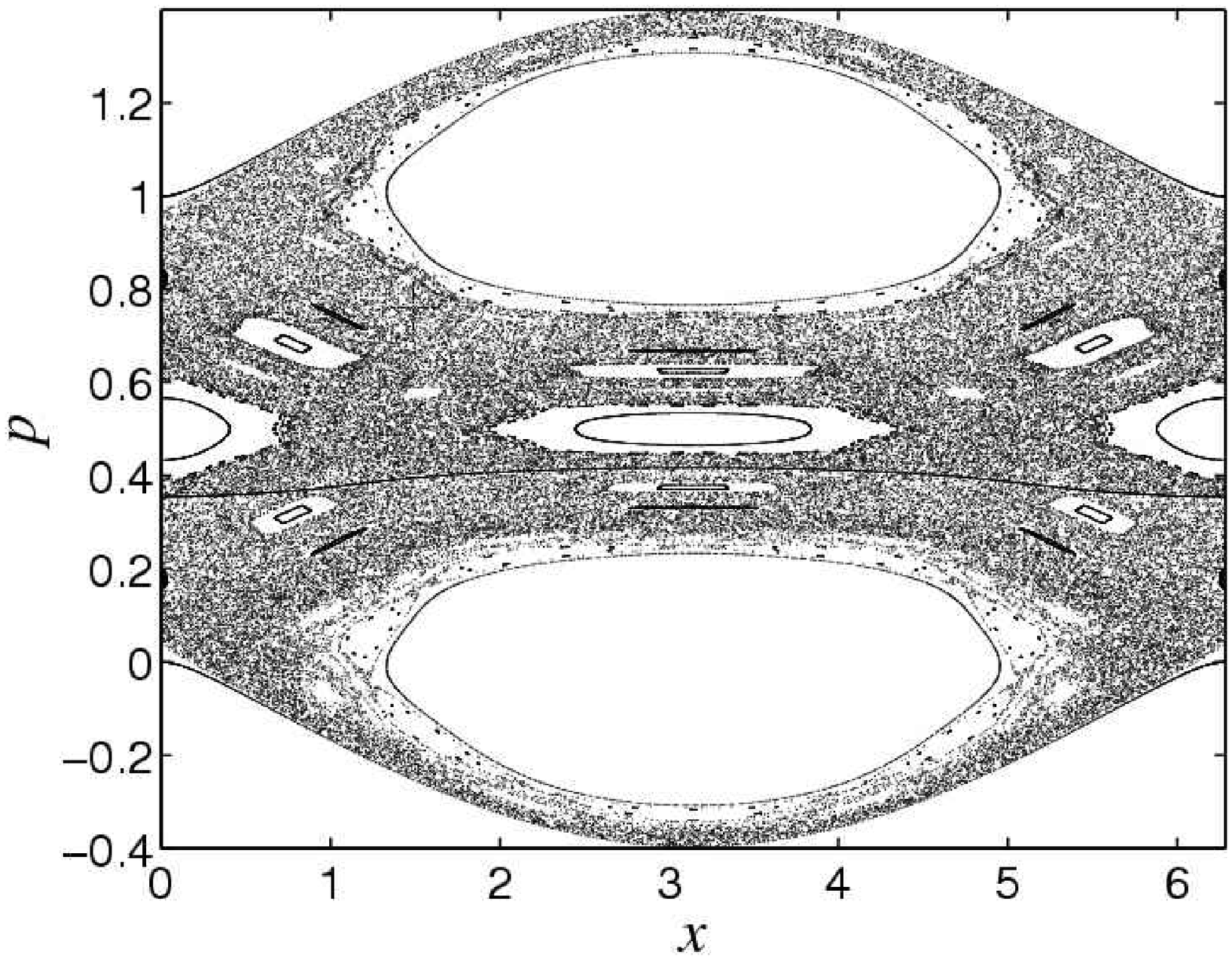,width=7.5cm,height=6.3cm}}
\put(3,3.9){\epsfig{file=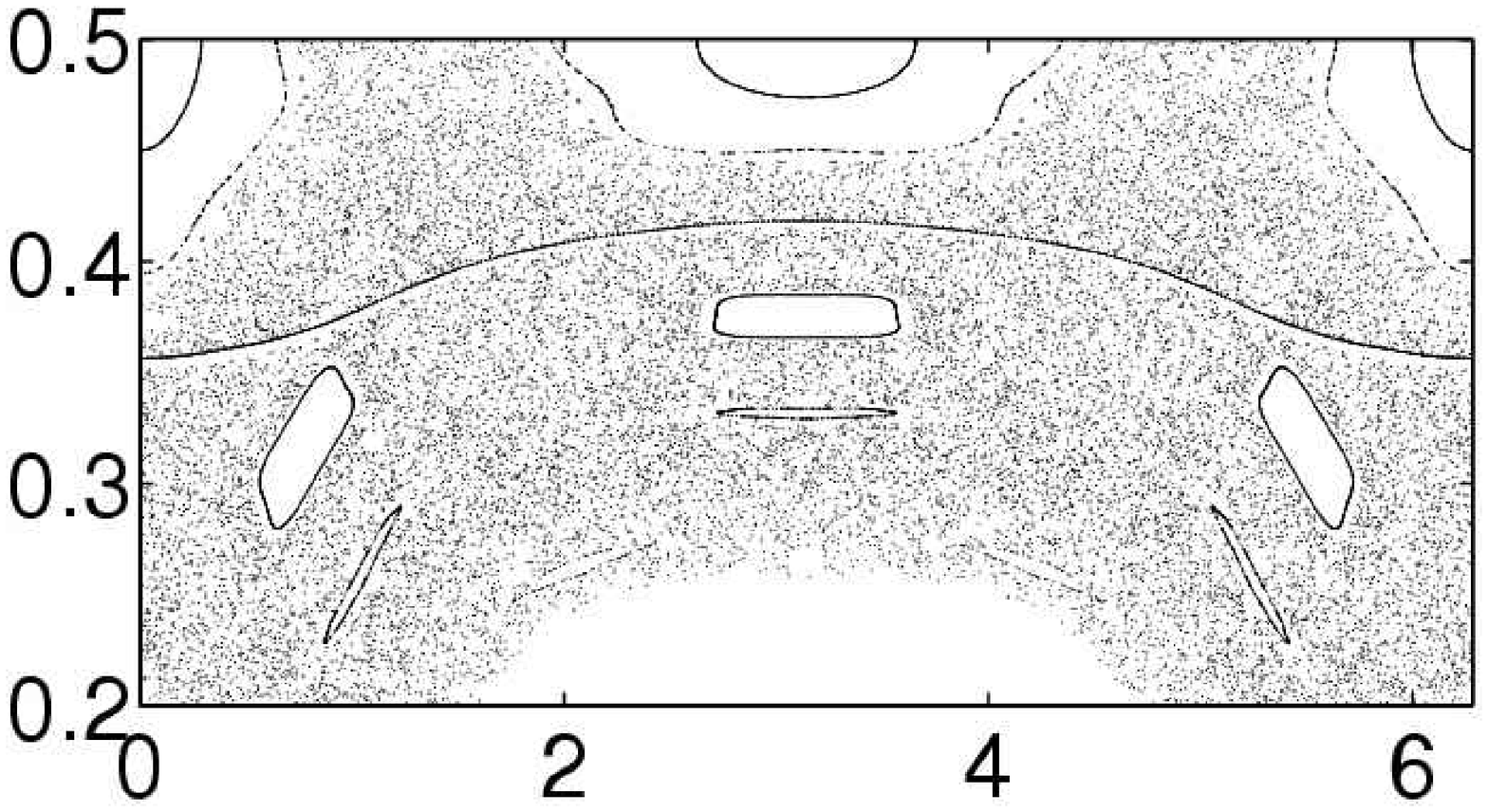,width=4.3cm,height=2.3cm}}
\end{picture}
\caption{Poincar\'e surface of section of Hamiltonian (\ref{eqn:fp}) with the approximate control term~(\ref{eqn:f2fpaL}) with $\varepsilon=0.034$ (enlargement in the inset) using the same initial conditions and same time of integration as in Fig.~\ref{fig1}.}
\label{fig2}
\end{figure}
The phase space of the controlled Hamiltonian appears to be very similar to the one of the uncontrolled Hamiltonian (see Fig.~\ref{fig2}). We notice that there is in addition an isolated invariant torus which is the one where the control term has been localized, i.e.\ its frequency is equal to $(3-\sqrt{5})/2$.\\

\begin{figure}
\epsfig{file=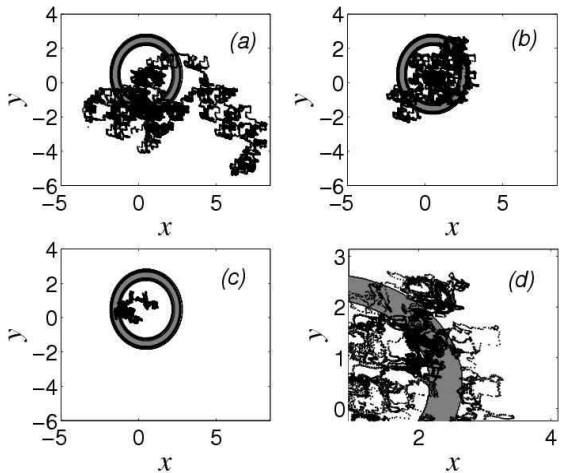,width=7.5cm,height=6.3cm}
\caption{Poincar\'e sections of ({\it a}) Hamiltonian (\ref{potential}), 
({\it b}) with localized control term (\ref{f_2_L}) 
and ({\it c}) with a full control term (\ref{f_2}) for $\varepsilon=0.7$ using the same initial conditions and same time of integration. An enlargement of 
({\it b}) is shown in ({\it d}).}
\label{fig3}
\end{figure}

\indent The second example comes from plasma physics. 
It aims at modeling chaotic 
${\bf E}\times{\bf B}$ drift motion. In the guiding center approximation, 
the equations of motion
of a charged particle in presence of a strong toroidal magnetic field
and of a nonstationary electric field are~:
$$
\frac{d}{dt}{x \choose y}=\frac{c}{B^2}{\bf E}(
x,y,t)\times {\bf B}= \frac{c}{B}{-\partial_y V (x,y,t)\choose
\partial_x V (x,y,t)},
$$
where $V$ is the electric potential, ${\bf E}=-{\bf\nabla} V$,
and ${\bf B}=B {\bf e}_z$. 
The spatial coordinates $x$ and $y$ where $(x,y)\in \mathbb{R}^2$ play 
the role of the canonically conjugate variables and the electric 
potential $V(x,y,t)$ is the Hamiltonian of the problem.
A model that reproduces the experimentally observed 
spectrum~\cite{anormal_exp} has been proposed in Ref.~\cite{marc88}. 
We consider the following explicit form of the electric potential
\begin{eqnarray}
V (x,y,t) &=&\frac{\varepsilon}{2\pi}\sum_{m,n=1\atop{n^2+m^2\le N^2}}^N
 \frac{1}{(n^2+m^2)^{3/2}}\times \nonumber 
\\ &&\sin \left[2\pi(nx + my-t) +
\varphi_{nm} \right],
\label{potential}
\end{eqnarray}
where the phases $\varphi_{nm}$ are chosen at random in order to model a 
turbulent electric potential. The control term has been computed in 
Ref.~\cite{guido1} from $H_0=E$ [i.e.\ a localized control 
using ${\bm \omega}=(1,0)$] where $E$ is the conjugate action to the 
angle $t \mbox{ mod } 2\pi$~:
\begin{eqnarray}
&&f_{2}(x,y)=
\frac{\varepsilon^2}{8\pi}\sum_{n_1,m_1\atop{n_2,m_2}} \frac{n_1 m_2 - n_2 m_1}{ 
 (n_1^2+m_1^2)^{3/2} (n_2^2+m_2^2)^{3/2}}  \nonumber\\
&&\times\sin \bigl[ 2\pi \bigl[ (n_1-n_2) x + (m_1-m_2) y\bigr] +
\varphi_{n_1 m_1} - \varphi_{n_2 m_2} \bigr] \nonumber.
\label{f_2}\\
\end{eqnarray}
In Ref.~\cite{guido1}, we showed that this global control term is 
able to strongly reduce
the chaotic diffusion of test particles provided that $\varepsilon \leq 1$, i.e. the 
trajectories of the controlled Hamiltonian $V + f_2$ diffuse much less than
the ones of the Hamiltonian $V$.\\
We now
localize the control term in a particular region of the phase space.
We choose this region to be a circular annulus of radius $r_0$ around a center 
$(x_0, y_0)$ in order to determine an area of confinement where the control can 
be applied on the circular border of an experimental apparatus.
We consider the distance from a point $(x,y)$ to the border 
of the apparatus (circle of radius $r_0$) : 
$d(x,y)=\mid r_0 - \sqrt{(x-x_0)^2 +(y-y_0)^2}\mid$.
The  localized control term  is given by
\begin{equation}
f_2^{(L)}(x,y)=f_2(x,y)\chi(d(x,y)),
\label{f_2_L}
\end{equation}
where  $\chi(d)=1$ if $d<\alpha$, 
$\chi(d)=0$ if $d> \beta$  and $\chi$ equals to a polynomial of degree three
for $d\in \left[\alpha, \beta\right]$ in order to ensure the derivability of 
the controlled Hamiltonian. This means that there is a region 
(the gray circular 
annulus in Fig.~\ref{fig3}) where the full control term is applied and
two border 
regions (in black) where the control term decreases to zero.
\begin{figure}
\epsfig{file=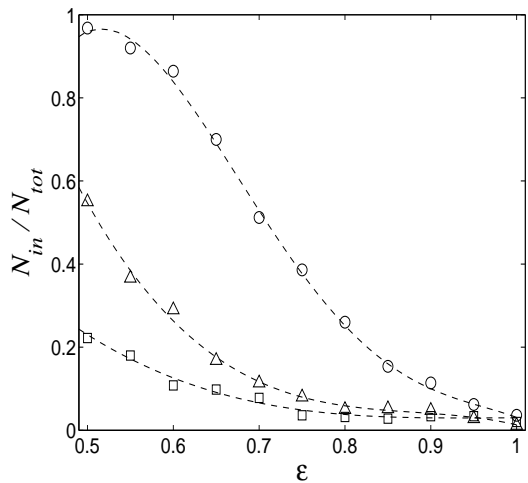,width=7cm,height=6.5cm}
\caption{Number of particles $N_{in}$ that remain inside 
the annulus during a time $t=30000$ by the total number $N_{tot}$ of particles 
versus the amplitude $\varepsilon$ of the 
electric potential (\ref{potential}) without control (open squares),
with  control term (\ref{f_2}) 
(open circles), and with the localized 
control term (\ref{f_2_L}) (up-triangles).}
\label{fig4}
\end{figure}
In Fig.~\ref{fig3}, we show some trajectories of  
particles in the cases without control term, with the full control term and
with the localized control term for Hamiltonian (\ref{potential}) with $\varepsilon=0.7$.
Whereas in the case without the control term a lot of particles 
exit the apparatus 
(the region $\Omega$ delimited by the circular annulus) and in the case 
with the full
control term the trajectories are very well confined. The localized control 
acts only on the border without modifying the chaotic
motion inside and outside
the region  $\Omega$ but catching the particles when they arrive close to the
border.
A measure of this effect is provided by the ratio of 
the number $N_{in}$ of particles that remain
inside the barrier after a given interval of time, 
divided by the total number $N_{tot}$ of particles.
In Fig.~\ref{fig4} the values of the ratio $N_{in}/N_{tot}$ are plotted versus 
the amplitude $\varepsilon$ of the electric potential (\ref{potential}).
It is shown that also this localized control is able to significantly reduce 
the escape of particles outside the barrier, although it is less efficient
than a full control which acts globally  on the system. 
We notice that the reduction of transport is not achieved by the 
creation of a KAM
torus as it is the case for the forced pendulum but by creating
a selected region of phase space where the system behaves much more 
regularly.\\
\indent A measure of the relative size of the control terms is given
by the electric energies, denoted by  
${\mathcal E}$, $e_2$ and $e^{(L)}_2$ associated with the electric 
potentials  $V$, $f_2$ and $f_2^{(L)}$, respectively. 
We define an electric energy 
${\mathcal E}=\langle~ |~{\bf E}~|^2~\rangle\cdot|\Omega|/8\pi$
where ${\bf E}(x,y,t)=-{\bf\nabla}V$ and $|\Omega|$ is the area where the
control acts.
For the potential (\ref{potential}) with $N=25$ and
for a circular annulus of thickness $0.7$, the relative amplitudes are:
$e_2^{(L)}/{\mathcal E}\approx 0.03\times \varepsilon^2$ and $e_2^{(L)}/e_2\approx 0.28$.
For instance, by choosing $\varepsilon=0.7$, $e_2^{(L)}$ is $1\%$ 
of $\mathcal E$. This comparison shows that the localized control of 
transport is energetically more efficient than a global control of chaos.\\
\indent In summary, we have provided a method of control of transport in Hamiltonian systems by designing a small control term, localized in phase space, that is able to create barriers of transport in situations of chaotic regime. Furthermore we remark that in view of practical applications, one main feature of our method of localized control is that only the partial knowledge of the potential on the designed support region is necessary.
Our approach opens the possibility of controlling the transport properties
of a physical system without altering the chaotic structure of its phase space.

\vfil

\end{document}